\newif\ifemulate
\emulatetrue

\ifemulate
  \documentclass{emulateapj}
  \usepackage{apjfonts}
\else
  \documentclass[12pt,preprint]{aastex}
\fi

\newcommand{\msun}{$M_\sun$}

\def\figa{
\begin{figure*}
\begin{center}
\leavevmode
\includegraphics[angle=0,width=16.75cm]{f1a.ps} \vspace{-1cm}
\includegraphics[angle=0,width=16.75cm]{f1b.ps} \vspace{-1cm}
\includegraphics[angle=0,width=16.75cm]{f1c.ps}
\includegraphics[angle=0,width=16.75cm]{f1d.ps}
\caption{Images of $z_{850}$-dropouts in the HUDF from 
HST/ACS ($i_{775},z_{850}$), NICMOS ($J_{110},H_{160}$), and
{\it Spitzer}/IRAC ($3.6\mu$m,$4.5\mu$m). All sources are 
undetected $(<2\sigma)$ at $i_{775}$ and bluer wavelengths, 
but we note that $z_{850}$-dropout galaxies can have some flux 
at $i_{775}$ due to incomplete absorption between
Lyman$-\alpha$ at rest-frame 1216\AA\ and the Lyman limit at 
912\AA. The top two candidates are clearly detected in the IRAC $3.6$ and 
$4.5\mu$m images, while two others are 
marginally detected. Flux contribution from nearby sources was subtracted 
in the Spitzer images. Each panel is $4\farcs1\times4\farcs1$
in size, or $\approx$21~kpc at $z=7$.}
\end{center}
\leavevmode
\vspace{-0.9cm}
\end{figure*}
}

\def\figb{
\begin{figure}
\includegraphics[angle=0,width=8.0cm]{f2.ps}
\caption{Examples of the deblending procedure in
the IRAC $3.6\mu$m image as performed separately for the
first ({\it top row}) and second ({\it bottom row}) 
epoch images from GOODS. 
We fit the original images ({\it left}) using the high resolution
NICMOS data to construct models for the nearby neighbors ({\it middle})
which are subtracted resulting in the cleaned images ({\it right}).
We recover consistent fluxes for both epochs
after subtraction. The panel sizes are $4\farcs1 \times 4\farcs1$. }
\end{figure}
}

\def\figc{
\begin{figure}
  \includegraphics[angle=0,width=8.0cm]{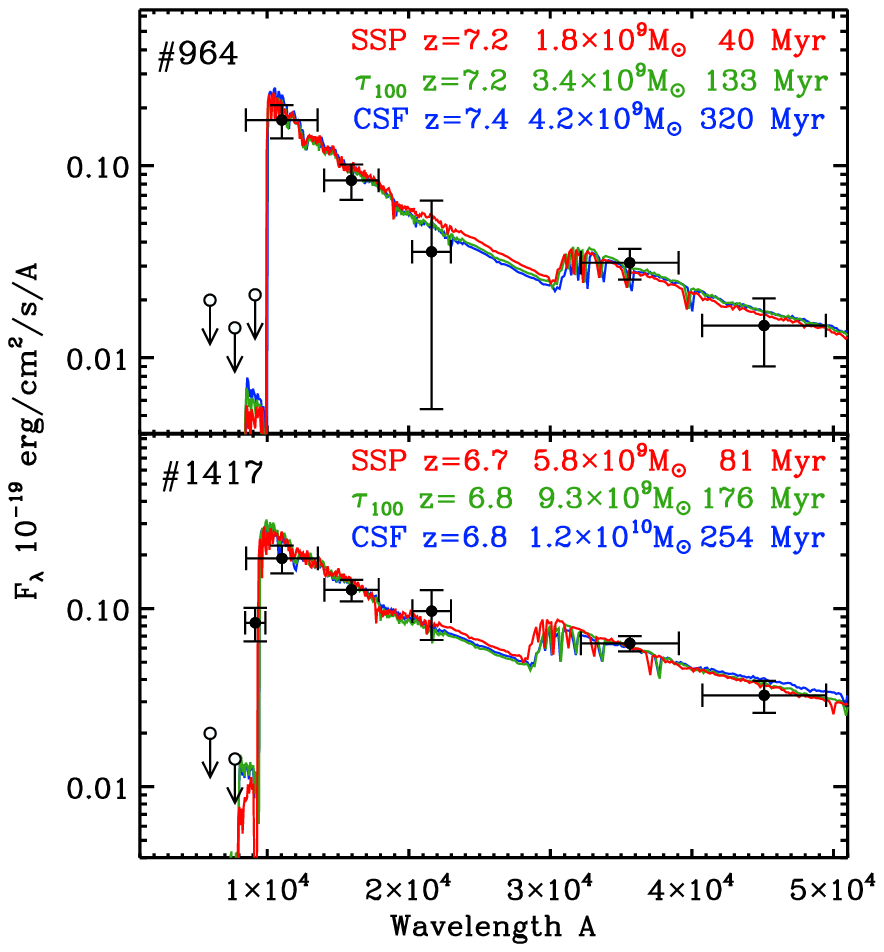}  
\caption{The observed SEDs of the IRAC-detected $z\approx$7 galaxies and the best-fit 
Solar metallicity models for each star formation history. The vertical
and horizontal bars mark the $1\sigma$ flux uncertainty and the
width of the passband. The arrows are $1\sigma$ upper limits.
The sources show evidence for a break between the near-infrared and mid-infrared fluxes,
which is well fit by a fairly evolved stellar population. The fits generally
require little reddening, although galaxy 1417 also allows for moderate 
amounts of reddening and fairly large star formation rates (up to $A_V=0.4$ and $SFR\sim25~$M$_\sun~$yr$^{-1}$).}
\end{figure}
\leavevmode
\vspace{-0.2cm}
}

\def\taba{
\begin{table} 
\begin{center}
\caption{Corrected Photometry of $z_{850}$-dropouts in the UDF.}
\begin{tabular}{ccccccc}
\tableline\tableline
ID & $z_{850}$ & $J_{110}$  & $H_{160}$ & $K_s$ & 3.6$\mu$m & 4.5$\mu$m \\ 
\tableline
\phs964 &  $>29.6$   & 26.8$\pm$0.2 & 26.8$\pm$0.2 & 27.0$\pm$0.9 &    26.1$\pm$0.2 & 26.4$\pm$0.5\\
1417 & 28.0$\pm$0.2 & 26.7$\pm$0.2 & 26.3$\pm$0.2 & 26.1$\pm$0.3 & 25.3$\pm$0.1 & 25.5$\pm$0.2\\
\phs950 &  $>29.6$  & 26.9$\pm$0.2 & 26.6$\pm$0.2 &     $>27.2$  &     27.0$\pm$0.5 & 26.4$\pm$0.5\\
1125 & 29.2$\pm$0.7 & 27.3$\pm$0.4 & 27.2$\pm$0.3 &  $>27.2$  & 26.9$\pm$0.4 &  $>27.2$  \\
\tableline
\end{tabular}
\tablecomments{Object IDs are from B04. 
All magnitudes are in the AB system. Optical/near-infrared 
fluxes were measured in 0\farcs9 diameter apertures and IRAC fluxes were measured in
2\farcs5 diameter apertures. We corrected the fluxes for missing light outside 
the aperture assuming stellar profiles. The corrections are  $5-20$\% in the 
optical/near-infrared and a factor 1.75, 1.81 for IRAC $3.6\mu$m, $4.5\mu$m,
respectively. We adjusted the NICMOS $J_{110}$ and $H_{160}$ zeropoints 
by -0.16 and -0.04 following \citet{Jo06}. In the improved NICMOS reduction (B06)
object 1417 falls within the $z_{850}$-dropout selection criteria, whereas 
in B04 it fell just outside. The upper limits are $1\sigma$. None of the 
galaxies are detected in $V_{606}, i_{775}$, and $5.8\mu$m to $1\sigma$ limits 
of 30.6, 30.5, and 25.1, respectively.}
\end{center}
\leavevmode
\vspace{-0.5cm}
\end{table}
}

\def\tabb{
\begin{table}
\begin{center}
\caption{Best-fit Model Parameters}
\setlength{\tabcolsep}{0.025in}
\scriptsize
\begin{tabular}{ccrrrcc}
\tableline\tableline
 & \multicolumn{6}{c}{Solar metallicity}  \\
ID & \phs\phs$z_{phot}$ & Mass  & Age$_w$ & $A_V$ & SFR & $\chi^2_{red}$ \\
& & $(10^9M_\sun)$ & (Myr) & & (\msun~yr$^{-1}$)  \\
\tableline
\multicolumn{6}{c}{Single Age Burst (SSP)} \\
\tableline
         964 & \ \ \  7.2 $^{+0.1}_{-0.2}$ &  1.8 $^{+0.7}_{-0.4}$ &    40 & \ \ \ \ \ 0.0 &  0.0 &  1.5  \\
        1417 & \ \ \  6.7 $^{+0.2}_{-0.2}$ &  5.8 $^{+0.9}_{-1.0}$ &    81 &  0.1 &  0.0 &  0.8  \\
         950 & \ \ \  7.2 $^{+0.1}_{-0.1}$ &  1.0 $^{+0.2}_{-0.1}$ &    25 &  0.0 &  0.0 &  1.4  \\
        1125 & \ \ \  7.0 $^{+0.2}_{-0.3}$ &  0.7 $^{+0.4}_{-0.1}$ &    32 &  0.0 &  0.0 &  0.9  \\
\tableline
\multicolumn{6}{c}{Exponentially Declining SFR (e-folding time $\tau=100$~Myr; $\tau_{100}$)}  \\
\tableline
         964 &  \ \ \ 7.2 $^{+0.1}_{-0.1}$ &  3.4 $^{+0.8}_{-0.7}$ &   133 &  0.0 &  5.2 &  1.2  \\
        1417 &  \ \ \ 6.8 $^{+0.1}_{-0.2}$ &  9.3 $^{+1.3}_{-2.2}$ &   176 &  0.2 & 10.7 &  0.7  \\
         950 &  \ \ \ 7.3 $^{+0.1}_{-0.2}$ &  0.6 $^{+0.5}_{-0.3}$ &    35 &  0.0 &  9.4 &  1.1  \\
        1125 &  \ \ \ 7.1 $^{+0.1}_{-0.2}$ &  0.9 $^{+0.8}_{-0.6}$ &    77 &  0.0 &  4.4 &  0.8  \\
\tableline
\multicolumn{6}{c}{Constant Star Formation (CSF)} \\
\tableline
         964 &\ \ \  7.4 $^{+0.1}_{-0.2}$ &  4.2 $^{+2.4}_{-2.1}$ &   320 &  0.0 &  7.3 &  1.3  \\
        1417 &\ \ \  6.8 $^{+0.2}_{-0.2}$ & 11.5 $^{+4.5}_{-3.6}$ &   254 &  0.4 & 25.5 &  0.9  \\
         950 &\ \ \  7.3 $^{+0.1}_{-0.2}$ &  0.7 $^{+0.6}_{-0.4}$ &    40 &  0.0 &  9.9 &  1.1  \\
        1125 &\ \ \  7.0 $^{+0.2}_{-0.2}$ &  1.0 $^{+1.3}_{-0.6}$ &   127 &  0.0 &  4.5 &  0.8  \\
\tableline
\end{tabular}
\tablecomments{
The best-fit parameters for Bruzual \& Charlot (2003) stellar population models 
and a \citet{Cal00} obscuration law. We omit the $B$ and $8\mu$m band from the 
fit as they are not deep enough to constrain the models, leaving 9 filters
($V_{606},i_{775},z_{850},J_{110},H_{160},K_{s},3.6\mu$m,
$4.5\mu$m, and 5.8$\mu$m), and 4 free parameters (redshift, stellar mass, age, 
and visual extinction $A_V$). We assume Solar metallicity and a 
Salpeter IMF from $0.1-100$~\msun; adopting a Chabrier 
IMF would result in similar ages but $\sim1.7 \times$ lower stellar masses.
Sub-Solar ($1/50~Z_\sun$) metallicities result in 30\% higher ages
and 20\% higher masses. We consider redshifts from 0 to 10 and $A_V$ from 0 to 2.
We explore 3 star formation histories (SFHs): SSP, $\tau_{100}$, and
CSF. The 68\% confidence intervals are obtained from
Monte-Carlo simulations and the reduced $\chi^2_{red}$ values assume 
5 degrees of freedom. We impose a minimum best-fit age of 25Myr 
to avoid unrealistically young solutions.
Finally, we weight the best-fit age with the SFH to better represent
the age of the stars comprising the bulk of the stellar mass (Age$_w$)
For CSF the correction is 0.5, for exponentially declining SFHs it is 
$[age-\tau+\tau exp(-age/\tau)]/[1-exp(-age/\tau)]$, while SSP
requires no correction \citep{Fo04}.}
\end{center}
\leavevmode
\vspace{-0.5cm}
\end{table}
}

\shorttitle{Spitzer IRAC confirmation of $z\approx7$ galaxies in the UDF}
\shortauthors{Labb\'e et al.}

\begin{document}


\title{{\it Spitzer} IRAC confirmation of $z_{850}$-dropout galaxies in the Hubble Ultra Deep Field: \\
       stellar masses and ages at $z\approx7$\altaffilmark{1}}


\author{Ivo labb\'e\altaffilmark{2,3},Rychard Bouwens\altaffilmark{4}, G.D. Illingworth\altaffilmark{4},
M. Franx\altaffilmark{5}}


\altaffiltext{1}{Based on observations with the {\em Spitzer Space Telescope}, which is 
operated by the Jet Propulsion Laboratory, California Institute of Technology under 
NASA contract 1407.  Support for this work was provided by NASA through contract 
125790 issued by JPL/Caltech. Based on observations with the
NASA/ESA {\em Hubble Space Telescope}, obtained at the Space
Telescope Science Institute which is operated by AURA, Inc.,
under NASA contract NAS5-26555. Based on service mode observations collected at 
the European Southern Observatory, Paranal, Chile (ESO Program 073.A-0764A).
}

\altaffiltext{2}{Carnegie Observatories, 813 Santa Barbara Street, Pasadena, CA 91101 [e-mail: {\tt ivo@ociw.edu}]}
\altaffiltext{3}{Carnegie fellow}
\altaffiltext{4}{Astronomy Department,University of California, Santa Cruz, CA 95064}
\altaffiltext{5}{Leiden Observatory, Postbus 9513, 2300 RA Leiden, Netherlands}

\begin{abstract}
Using {\it Spitzer} IRAC mid-infrared imaging from the Great Observatories Origins 
Deep Survey, we study $z_{850}$-dropout 
sources in the Hubble Ultra Deep Field. After carefully removing
contaminating flux from foreground sources, we clearly detect 
two $z_{850}$-dropouts at $3.6\mu$m and $4.5\mu$m, while two others 
are marginally detected. 
The mid-infrared fluxes strongly support their interpretation as galaxies at 
$z\approx7$, seen when the Universe was only 750~Myr old. 
The IRAC observations allow us
for the first time  to constrain the rest-frame optical colors, stellar masses, 
and ages of the highest redshift galaxies.
Fitting stellar population models to the spectral energy 
distributions, we find photometric 
redshifts in the range $6.7-7.4$, rest-frame colors $U-V=0.2-0.4$, $V$-band
luminosities $L_V=0.6-3 \times 10^{10}~L_\sun$, 
stellar masses $1-10 \times 10^{9}$~\msun,
stellar ages $50-200$~Myr, star formation
rates up to $\sim25$~\msun~yr$^{-1}$, and low reddening $A_V<0.4$.
Overall, the $z=7$ galaxies appear substantially less 
massive and evolved than Lyman break galaxies or Distant
Red Galaxies at $z=2-3$, but fairly similar to recently 
identified systems at $z=5-6$. 
The stellar mass density inferred from our $z=7$ sample is 
$\rho_* = 1.6^{+1.6}_{-0.8}\times 10^{6}$~\msun Mpc$^{-3}$ (to 0.3$L^*_{z=3}$), in 
apparent agreement with recent cosmological hydrodynamic simulations,
but we note that incompleteness and sample variance may introduce 
larger uncertainties.
The ages of the two most massive galaxies suggest they 
formed at $z\gtrsim8$, during the era of cosmic reionization,
but the star formation rate density derived from their stellar masses
and ages is not nearly sufficient to reionize the universe. 
The simplest explanation for this deficiency is that lower-mass galaxies beyond our
 detection limit reionized the universe.

\end{abstract}




\section{Introduction}
Observations of massive galaxies at high redshift
 with the {\it Hubble Space Telescope} and the 
{\it Spitzer Space Telescope} are revolutionizing 
our knowledge of the early formation history of stars and galaxies. 
Blue star forming galaxies at $z=2-3$ with stellar masses
 $10^{10}-10^{11}$\msun\ are routinely identified 
 from optical imaging (Steidel et al. 1996a,b,2004),
and have been studied in detail 
\citep[e.g.,][Shapley et al. 2001,2005]{PDF01}, while
near-infrared surveys at $z=2-3$ have uncovered substantial numbers of 
redder, more evolved galaxies with larger stellar masses
$>10^{11}$\msun\  \citep{Fr03,Ya04,Da05}. Some of these red galaxies 
appear to have stellar ages $> 1.5$ Gyr,
implying that they formed most of their stars before $z\sim5$ \citep{La05},
and suggesting that massive galaxies should exist well beyond these redshifts.
Direct detection of such galaxies would place strong 
constraints on galaxy formation models (e.g., \citealt{So01,Na05}).

Tantalizingly, the most recent surveys with the Advanced Camera for Surveys 
 (ACS) and the Near-Infrared Camera and Multiobject Spectrograph (NICMOS) 
on the {\it Hubble Space Telescope} have identified 
sources at $z=5-6.5$ with fairly evolved stars and
stellar masses of $1-4 \times 10^{10}$\msun\ \citep[e.g.,][]{Ya05,Ey05,Do05} or perhaps 
more \citep{Mo05}. Critical for these results was access to the rest-frame 
wavelengths longward of the Balmer/4000 \AA\ break offered by the 
InfraRed Array Camera (IRAC; \citealt{Fa04}) on {\it Spitzer}. 
Without mid-infrared photometry to very faint magnitudes the
stellar ages and masses of $z\gtrsim5$  galaxies are  poorly constrained.

In this Letter, we extend mass estimates to $z=7-8$
by analyzing the mid-infrared fluxes of 6 $z_{850}$-dropout
candidates found in the Hubble Ultra Deep Field (UDF)
by Bouwens et al. (2004, hereafter B04). These candidates 
were selected from exceptionally deep optical ACS 
(Beckwith et al. in prep) and near-infrared NICMOS 
imaging \citep{Th05}, and when combined with the 
ultradeep IRAC data available there, offer us an 
ideal opportunity to verify their reality and to study
their stellar populations. The stellar masses and ages of $z_{850}$-dropout galaxies 
would provide us with the first direct look at galaxy formation at $z\gtrsim7$,
building on the comprehensive $z\sim6$ study (Bouwens et al. 2006b).
Where necessary, we assume an 
$\Omega_M=0.3, \Omega_\Lambda=0.7,$ cosmology with 
$H_0=70$~km~s$^{-1}$Mpc$^{-1}$.
Magnitudes are expressed in the AB photometric system.  

\ifemulate\figa\fi

\section{Observations}
The Great Observatories Origins Deep Survey (GOODS; Dickinson et al., in prep) 
observed the UDF with IRAC in two epochs, 
each time integrating for $\approx23.3$ hours in the 
$3.6, 4.5, 5.8$, and $8.0\mu$m channels.\footnote{This paper uses 
data release DR3 of epoch 1 and data release DR2 of epoch 2, available 
from \url{http://data.spitzer.caltech.edu/popular/goods/}}
We estimate limiting depths
in the combined IRAC images by measuring 
the effective flux variation in random apertures on 
empty background regions. The
limits for point sources are 27.7, 27.2, 25.1, and
24.9 in the four channels (1$\sigma$, total, 2\farcs5 diameter
aperture).
We supplement the observations with deep $K_s$-band
data from the {\it Very Large Telescope} and {\it Magellan}
(Labb\'e et al. in prep) and we use an independent reduction of 
UDF NICMOS data with improved noise properties and fewer
artifacts \citep[][hereafter B06]{Bo06}. The new NICMOS images revealed that 
2 of the original 6 $z_{850}$-dropouts were electronic ghosts 
of nearby bright stars; hence we removed them from the sample.

Matching ACS/NICMOS and IRAC photometry is challenging
because of the much larger size and extended wings of the 
IRAC point-spread functions (PSFs) resulting in
flux contamination by nearby foreground sources. Visual
inspection shows that 2 $z_{850}$-dropouts are substantially
blended and all 4 are likely to contain at least some flux 
from nearby objects.
We have developed a technique to robustly subtract the 
contaminating flux (Labb\'e et al., in prep). 
Briefly, we first detect sources
with SExtractor \citep{BA96} in a summed NICMOS $J_{110} + H_{160}$ image
to determine the light distributions at high resolution using the pixels 
in the ``segmentation'' maps. We then convolve these template
images individually with a carefully constructed kernel to match it 
to the IRAC PSF.  Third, we fit all detected sources, including  
neighbors, simultaneously to the registered IRAC image, leaving 
only the flux scalings as free parameters. 
Finally, we subtract the best-fit images of all neighboring sources. 

\ifemulate\taba\fi

After cleaning the IRAC images, we performed conventional
aperture photometry in $3.6,4.5,5.8$, and $8.0\mu$m bands
in 2\farcs5 diameter apertures. 
Photometry in the ACS $B_{435},V_{606},i_{775},z_{850}$, 
NICMOS $J_{110},H_{160}$, and  $K_s$ bands was done in 
0\farcs9 diameter apertures and we obtained
magnitudes and limits consistent with B04.
We summarize the photometry in Table~1.

\section{Mid-Infrared Fluxes of $z_{850}$-dropout sources}
Figure 1 shows the {\it HST}/ACS+NICMOS and the combined
 {\it Spitzer}/IRAC images of the $z_{850}$ dropouts. 
 Two objects (ID 964 and 1417) are unambiguously detected 
in $3.6\mu$m ($5-8 \sigma$) and in the slightly shallower 
$4.5\mu$m ($2-3 \sigma)$. Two others 
(ID 950 and 1125) are only marginally detected, but 
are probably real as the sum of their $3.6$ and $4.5\mu$m 
images reveals a visible source. Unfortunately, the IRAC 
observations are not deep enough to definitively 
confirm or reject the reality of the undetected sources.
None of the candidates are visible at $5.8$ and $8.0\mu$m.
To evaluate the robustness of the deblending photometry, we performed
the procedure independently on the first and second epoch 
IRAC data (see Figure~2). Reassuringly, we measure consistent fluxes 
and we detect the brightest,
most promising sources 964 and 1417 in each dataset individually.
 
The $3.6\mu$m magnitudes are faint, ranging from 25.3 to 27.0, 
with $H_{160}-3.6\mu$m colors in the range -0.4 to 1.0. The IRAC detected
objects 964 and 1417 are the reddest, showing a 
factor of $\sim2$ rise in $f_\nu$ flux densities between $H_{160}$ 
and $3.6\mu$m, while the SEDs are flatter at $1.1-2.2\mu$m and 
$3.6-4.5\mu$m (see Table~1). The rise at $3.6\mu$m is similar in 
strength to what is found in spectroscopically confirmed galaxies 
at $z=6$ \citep{Ya05,Ey05}, and suggests the presence of 
a substantial redshifted Balmer break, indicative of evolved 
stellar populations.  

\ifemulate\figb\fi

\section{Stellar Populations of $z\approx7$ galaxies}
We fit stellar population synthesis models of \citet{BC03} 
and a \citet{Cal00} obscuration law to the broadband fluxes to 
constrain the stellar populations. The models assume Solar metallicity
and a Salpeter initial mass function (IMF) between 0.1 and 100~\msun. 
We explore three different star formation histories (SFHs): 
a single age burst (SSP), an exponentially declining star 
formation rate (SFR) with a timescale of 100~Myr ($\tau_{100}$), and constant 
star formation (CSF). We leave the redshift, mass, age, 
and extinction as free parameters.

We find acceptable fits for all sources (see Table~2 and Figure~3) and 
obtain confidence intervals on the parameters with 
Monte Carlo simulations. 
The best-fit redshifts vary from 6.7 to 7.4 and
most Monte Carlo solutions are in a narrow range 
around the best fit indicating that the redshift is well constrained.
Old stellar populations at $z\sim1$ fit the data
poorly as they do not reproduce the strong
break across the $i_{775},z_{850}$, and 
$J_{110}$ bands and the blue near-infrared continuum. 
Even so, we note that 3\%  and 11\% of the solutions 
for object 1417 and 1125, allowed a redshift 
of $z\sim1$ when the random flux variations 
``conspired'' to suppress the break. The best-fit models have
average rest-frame optical colors of $U-V=0.4~(0.2)$ and 
$V-$band luminosities $L_V=2.3(1.0) \times 10^{10}L_\sun$ for the IRAC-detected 
(undetected) sources.

\ifemulate\tabb\fi
\ifemulate\figc\fi

Determinations of the stellar population age and mass
 depend on the assumed star formation history and 
 metallicity. For the whole sample, $\tau_{100}$
models fit the best, with ages of $40-180$~Myr and instantaneous 
SFRs of $4-11$~\msun~yr$^{-1}$.
Converting the rest-frame $1500~$\AA\ luminosities directly 
into (absorption-corrected) SFRs \citep{Ma98} results 
in similar values. Nevertheless, the degeneracy between age and dust
prevents us from placing firm limits on the ages and SFRs.
The stellar masses are generally better constrained. As expected,
the IRAC-detected galaxies 964 ($z_{ph}=7.4$) and 1417 ($z_{ph}=6.8$)
are the reddest, most massive, and oldest in the sample. 
The average uncertainties on the masses are approximately 
a factor of $2-3$. 

The extreme SSP and CSF models set lower and 
upper boundaries to the masses, ages, and SFRs, whereas
assuming sub-Solar metallicities
($1/50~Z_{\sun}$) instead of Solar results in 30\%
higher ages and 20\% higher masses.
Because all SFHs and metallicities provide equally acceptable 
fits to the data, we will hereafter adopt the mean of the 
SSP and CSF models and both metallicities 
as our fiducial values. We then find ages of $50-200$~Myr, 
masses of $1-10 \times 10^9$~\msun, low reddening $A_V<0.4$, 
and star formation rates of $3-12$~\msun~yr$^{-1}$.

\section{Discussion}

Using the GOODS dataset (Dickinson et al. in prep),
we have estimated and analyzed the {\it Spitzer}/IRAC 
mid-infrared fluxes of 4 $z_{850}$-dropouts
candidates, which were identified in the UDF by B04 
and remeasured more accurately in B06. The sources
are rare, with a surface density of 0.7~arcmin$^{-2}$, 
and very faint, with observed magnitudes of $H_{160}=26-27$ 
and $3.6\mu=25-27$, placing them well beyond the 
spectroscopic capabilities of current telescopes, but 
in reach of future facilities such as JWST and ALMA. 
IRAC directly confirms the reality of two sources, while 
two others are marginally detected. Modeling of the broadband 
fluxes strongly supports their interpretation 
as $z\approx7$ galaxies with substantial stellar masses 
$1-10 \times 10^9$ and ages $50-200$~Myr.

Using the redshift selection function for the
$z_{850}$-dropout sample (B04), we obtain an effective volume of 
9000 Mpc$^{3}$ to $0.3L^*_{z=3}$ and we infer a stellar mass 
density of $\rho_* = 1.6^{+1.6}_{-0.8}\times 10^{6}$~\msun Mpc$^{-3}$.
Comparing to the stellar mass density at lower redshifts, computed using 
similar techniques to similar luminosities, we find a continuing decrease where 
the density at $z\approx7$ is 95\%, 22\% of that at $z=6,5$  \citep{Ya06,Sta06}.
Recent Smooth Particle
Hydrodynamics (SPH) simulations in a $\Lambda$CDM universe 
 predict stellar mass- and number densities for massive ($>1.8 \times 10^{9}$M$_\sun$)
galaxies of $0.9 \times 10^{6}$~\msun~Mpc$^{-3}$ and 
$2.5 \times 10^{-4}$~Mpc$^{-3}$ (SPH G6 run, \citealt{Na05}),
remarkably close to the minimum estimates for our sample  
($0.7 \times 10^{6}$~\msun Mpc$^{-3}$ and $2.2 \times 10^{-4}$~Mpc$^{-3}$ to the same mass limit).
However, we cannot exclude additional stellar mass residing
in massive non-starforming or dust-enshrouded galaxies, 
which the $z_{850}$-dropout criteria would have missed. 

The ages of the two most massive (IRAC-detected) $z_{850}$-dropouts 
suggest that the bulk of their stellar mass formed at 
even higher redshifts $z\gtrsim8$, during the epoch 
of cosmic reionization \citep{Sp06}. 
We can place a simple upper limit on the contribution
of high-mass galaxies to reionization,
by calculating the maximum SFR densities implied by the 
observed stellar masses and ages for these objects.
Taking the maximum masses and dividing it by 
the minimum ages ($\sim60$Myr), we infer a substantial 
SFR density 0.04~M$_\sun$yr$^{-1}$~Mpc$^{-3}$. This is higher than calculated 
directly from the rest-frame UV luminosities at $z=6-7$  
to the same limits \citep{Bo06b}, but  still more than $3$ times too small to reionize 
the universe at the lowest probable redshift $z_{reion}=8.6$ \citep{Sp06} 
for canonical assumptions \citep[][see also Yan et al. 2006]{Ma99,Bo06b}. 

The simplest explanation for the lack of ionizing
photons provided by high-mass galaxies at $z=7$
is that low-mass galaxies beyond our current
detection limits were primarily responsible for reionization
\citep{LB03,YW04,Bo06b}.
Other possibilities are that massive galaxies are missing 
from current surveys and are unaccounted for in the models,
or that the observed galaxies had a top heavy IMF, which would 
increase the ionizing efficiency per unit stellar mass.
Obviously, incompleteness, sample variance, and 
large scale structure dominate the uncertainties in our
results. Larger, very deep near-infrared surveys from the ground
and from space will address this issue in more detail.

\acknowledgments
\vskip -0.2cm
We are most grateful for the efforts of the GOODS team who provided the
data used in this analysis. We thank Ken Nagamine for providing his
numerical simulations and the referee for helpful comments. 
IL is supported by a fellowship from the Carnegie Institution
of Washington. RJB and GDI acknowledge support from NASA grant HST-GO09803.05-A 
and NAG5-7697.



\ifemulate\else
  \figa
  \figb
  \figc
  \clearpage
  \taba
  \clearpage
  \tabb  
\fi


\begin{thebibliography}{}
\bibitem[Bertin \& Arnouts(1996)]{BA96} Bertin, E.~\& Arnouts, S.\ 1996, \aaps, 117, 393
\bibitem[Bouwens et al.(2004)]{Bo04} Bouwens, R.~J., et al.\ 2004, \apjl, 616, L79  (B04)
\bibitem[Bouwens \& Illingworth(2006)]{Bo06} Bouwens, R.~J.,  \& Illingworth, G.~D., \ 2006, Nature, in press \\ astro-ph/0607087 (B06)
\bibitem[Bouwens et al.(2006b)]{Bo06b} Bouwens et al.\ 2006b, \apj, in press, astro-ph/0509641
\bibitem[Bruzual \& Charlot(2003)]{BC03}  Bruzual, G.~\& Charlot, S.\ 2003, \mnras, 344, 1000 (BC03)
\bibitem[Calzetti et al.(2000)]{Cal00} Calzetti, D., et al.\ 2000, \apj, 533, 682 
\bibitem[Daddi et al.(2005)]{Da05} Daddi, E., et al.\ 2005, \apj, 626, 680 
\bibitem[Dow-Hygelund et al.(2005)]{Do05} Dow-Hygelund, C.~C., et al.\ 2005, \apjl, 630, L137 
\bibitem[de Jong et al.(2006)]{Jo06} de Jong, R.~S., et al.\ 2006, NICMOS ISR 2006-001
\bibitem[Eyles et al.(2005)]{Ey05} Eyles, L.~P., et al.\ 2005, \mnras, 364, 443 
\bibitem[Fazio et al.(2004)]{Fa04} Fazio, G.~G., et al.\  2004, \apjs, 154, 10 
\bibitem[F\"orster Schreiber et al.(2004)]{Fo04} F\"orster Schreiber, N.~M., et al.\ 2004, \apj, 616, 40 
\bibitem[Franx et al.(2003)]{Fr03} Franx, M.~et al.\ 2003, \apjl, 587, L79
\bibitem[Labb{\'e} et al.(2005)]{La05} Labb{\'e}, I., et al.\ 2005, \apjl, 624, L81 
\bibitem[Lehnert \& Bremer(2003)]{LB03} Lehnert, M.~D., \& Bremer, M.\ 2003, \apj, 593, 630 
\bibitem[Madau et al.(1998)]{Ma98} Madau, P., Pozzetti, L., \& Dickinson, M.\ 1998, \apj, 498, 106 
\bibitem[Madau et al.(1999)]{Ma99} Madau, P., Haardt, F., \& Rees, M.~J.\ 1999, \apj, 514, 648 
\bibitem[Mobasher et al.(2005)]{Mo05} Mobasher, B., et al.\ 2005, \apj, 635, 832 
\bibitem[Nagamine et al.(2005)]{Na05} Nagamine, K., et al.\ 2005, \apj, 627, 608 
\bibitem[Papovich, Dickinson, \& Ferguson(2001)]{PDF01} Papovich, C., Dickinson, M., \& Ferguson, H.~C.\ 2001, \apj, 559, 620
\bibitem[Stark et al.(2006)]{Sta06} Stark et al., 2006, submitted to ApJ, astro-ph/0604250 
\bibitem[Steidel et al.(1996a)]{St96a} Steidel, C.~C., et al. \ 1996, \aj, 112, 352
\bibitem[Steidel et al.(1996b)]{St96b} Steidel, C. C., et al. 1996b, \apj, 462, L17
\bibitem[Steidel et al.(2004)]{St04} Steidel, C.~C., et al.\ 2004, \apj, 604, 534 
\bibitem[Somerville et al.(2001)]{So01} Somerville, R.~S., Primack, J.~R., \& Faber, S.~M.\ 2001, \mnras, 320, 504 
\bibitem[Shapley et al.(2001)]{Sh01} Shapley, A.~E., et al.\ 2001, \apj, 562, 95 
\bibitem[Shapley et al.(2005)]{Sh05} Shapley, A.~E., et al.\ 2005, \apj, 626, 698 
\bibitem[Spergel et al.(2006)]{Sp06} Spergel et al., 2006, submitted to \apj
\bibitem[Thompson et al.(2005)]{Th05} Thompson, R.~I., et al.\ 2005, \aj, 130, 1 
\bibitem[Yan et al.(2004)]{Ya04} Yan, H., et al.\ 2004, \apj, 616, 63 
\bibitem[Yan \& Windhorst(2004)]{YW04} Yan, H., \& Windhorst, R.~A.\ 2004, \apjl, 600, L1 
\bibitem[Yan et al.(2005)]{Ya05} Yan, H., et al.\ 2005, \apj, 634, 109 
\bibitem[Yan et al.(2006)]{Ya06} Yan, H., et al.\ 2006, \apj, in press
\end{thebibliography}
\end{document}